# Evolution of Structure and Magnetic Properties of $Cu_2MnBO_5$ under Partial $Mn^{3+} \rightarrow Fe^{3+}$ Substitution

Moshkina E.M.[1,2], Platunov M.S.[1], Seryotkin Yu.V.[3,4], Bovina A.F.[1], Eremin E.V.[1], Sofronova S.N.[1], Bezmaternykh L.N.[1]

[1] Kirensky Institute of Physics, Federal Research Center KSC SB RAS, Krasnoyarsk, 660036 Russia
[2] Siberian State University of Science and Techlogies, Krasnoyarsk, 660014 Russia
[3] V.S. Sobolev Institute of Geology and Mineralogy, SB RAS, 630090 Novosibirsk, Russia
[4] Novosibirsk State University, 630090 Novosibirsk, Russia

Single crystals of Fe-substituted $Cu_2Mn_{1-x}Fe_xBO_5$ ludwigites have been synthesized using flux technique (x=0.2, 0.4, 0.5 – in the initial flux system). Structural properties of the synthesized compounds were studied by the single crystal and powder X-ray diffraction analysis. Obtained results were analyzed in the relationship with parent compound $Cu_2MnBO_5$. It was revealed that the type of monoclinic distortions of Fe-substituted ludwigites is different from the structure of $Cu_2MnBO_5$. The real cation composition and local structure of $Cu_2Mn_{1-x}Fe_xBO_5$ ludwigites were studied using XANES and EXAFS techniques, respectively. Analysis of field and thermal dependencies of magnetization showed a strong dependence of the magnetic properties of these ludwigites on *x* with changing the type of magnetic ordering.

## I. Introduction

Many dielectric magnetic materials already used in different technologies or quite perspective for future applications are based on $Fe^{3+}$ or $Mn^{3+}$ [1-5]. So an understanding of micromechanisms of the effects related to magnetic properties of compounds containing $Mn^{3+}$ or $Fe^{3+}$, or $Mn^{3+}$ and $Fe^{3+}$ cation is the problem of high scientific importance. Compounds containing trivalent cations of $Mn^{3+}$ in octahedral coordination are usually have other properties than the identical compounds containing $Fe^{3+}$ cations [6, 7]. There are many cases where the cause of such difference is an symmetry of the octahedra provided by different configurations of electron outer shell: for $Fe^{3+}$ the symmetric coordination is preferable, but for $Mn^{3+}$ the tetragonal distortion of the octahedron is more preferable due to the Jahn-Teller effect [6].

There are a lot of compounds with simultaneous content of the manganese and iron cations, which form series of solid solutions in some cases (for ex. [6, 8]). Despite the ordered magnetic state of pure either manganese or iron compounds, solid solutions do not possess the long range order as a rule [9]. For many cases such magnetic behavior is a consequence of structure disorder: ions $Mn^{3+}$ and $Fe^{3+}$ in solid solutions are distributed over crystallographic positions statistically.

To date there is a high scientific interest to quasi-low-dimensional oxyborates of transitional metals which are classified as strongly correlated systems. One of the most exciting parts of this family is oxyborates with ludwigite structure. Ludwigites have a large variety of cation substitutions and high dependence of the magnetic properties even on the small deviation of the composition. The quasi-low-dimensionality of ludwigites lies the presence of zig-zag walls consisting of the connected metal-oxygen octahedra and separated by boron-oxygen triangles (Figure 1). Ludwigite unit cell contains four formula units and four nonequivalent crystallographic positions of transitional metals having different valence states (di- and trivalent or di- and tetravalent). Ludwigite structure is characterized by the large number of triangular groups formed by metal cations, which also could lead to the occurrence of frustrations [10].

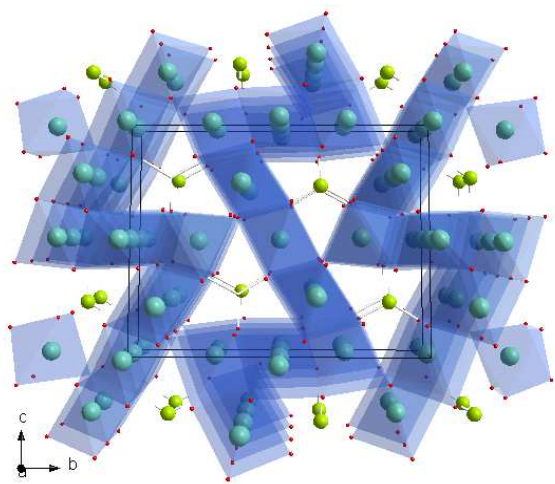

Figure 1. Ludwigite structure

Recently a new member of ludwigite family $Cu_2MnBO_5$ has been synthesized [10-12]. And to date, this compound is the object of intensive scientific research [10-13]. A special attention to this ludwigite is caused by the unusual behavior of the magnetic parameters in comparison with the other ludwigites – ferrimagnetic type of magnetic ordering at quite high temperature T=92 K relatively the isostructural analogues, and a quite large magnetic moment both along the *a* axis and along the *bc* plane. In addition, in $Cu_2MnBO_5$ there are two Jahn-Teller ions affecting the crystallographic and magnetic structure to a large extent – directions of $Cu^{2+}$ and $Mn^{3+}$ magnetic moments do not coincide with the directions of the principal crystallographic axes. Magnetic structure of $Cu_2MnBO_5$ is non-centrosymmetric, and despite quite large macroscopic magnetic moment, in agreement with powder neutron diffraction data, one of the four crystallographic positions occupied by Cu ions is not fully ordered [10]. To date the ludwigite $Cu_2MnBO_5$ is the first and so far only one heterometallic ludwigite with experimentally defined magnetic structure.

The ludwigite $Cu_2FeBO_5$ demonstrates the absolutely different behavior of magnetic properties than Mn-containing analogue. Moreover, there are several papers on this compound containing different results on the magnetization behavior [14, 15]. In agreement with [14], $Cu_2FeBO_5$ is antiferromagnet at low temperatures via three sequential phase transitions: at $T_f$=63K – transition to spin glass state of the iron subsystem, at $T_{N1}$=38K – antiferromagnetic ordering of the copper subsystem, at $T_{N2}$=20K – antiferromagnetic ordering of the iron subsystem. In agreement with [15] $Cu_2FeBO_5$ is also antiferromagnet at low temperature but there are only one phase transition at $T_N$=32 K. The comparison of the Mössbauer spectra studying results of [14] and [15] gives the different cation distributions over four metallic positions, it could indicate the dependence of the Fe-distribution on the synthesis technique.

Thus the study of the synthesis and magnetic properties of solid solutions $Cu_2Mn_{1-x}Fe_xBO_5$ (0<x<1) is an important problem which could help in understanding the microscopic nature of the magnetic behavior and difference of physical properties of Mn- and Fe- containing compounds. From the point of view of exchange interactions the important role in magnetic ordering belongs to balance between 90° Fe-O-Fe (Mn-O-Mn) and Fe(Mn)-O-Cu superexchange interactions and 180° Fe(Mn)-O-Cu exchange interactions. In agreement with

Goodenough–Kanamouri rules all of these exchange interactions are antiferromagnetic. The magnetic order is defined by the strongest of them.

In the present work we report flux synthesis conditions (II), structure characterization, carried out using single crystal and powder X-ray analyzes (III-a) and EXAFS/XANES techniques (III-b) as well as macroscopic magnetic properties analysis (IV) of $Cu_2Mn_{1-x}Fe_xBO_5$ ludwigites in comparison with unsubstituted $Cu_2MnBO_5$ ludwigite.

## II. Crystal Growth

Single crystals of $Cu_2Mn_{1-x}Fe_xBO_5$ were synthesized by the flux method. In the initial flux system the stoichiometric mixture $\frac{(1-x)}{2}Mn_2O_3:\frac{x}{2}Fe_2O_3:2CuO:0.5B_2O_3$ had been dissolved in the mixture $Bi_2Mo_3O_{12}:pB_2O_3:qNa_2O$ with concentrations $n$. The parameters $x$, $q$, $p$, $n$, $T_{sat}$ are shown in Table 1. The parameter $q$ (content of the sodium oxide $Na_2O$) is increasing with the increase of the iron content. The total sodium oxide content was calculated via the formula:

$$2NaMn^{3+}_{1-x}Fe_xO_2 \rightarrow Na_2O+(1-x)Mn_2O_3+xFe_2O_3$$

The fluxes in masses of 77-95 g were prepared from crystal-forming oxides $Mn_2O_3$, $Fe_2O_3$, $B_2O_3$ and CuO and solvent $Bi_2Mo_3O_{12}$ in combination with sodium carbonate $Na_2CO_3$ at the temperature T=1100°C in a platinum crucible with the volume V=100 cm$^3$ by sequential melting of powder mixtures, first $Bi_2Mo_3O_{12}$ and $B_2O_3$, then $Na_2CO_3$ was added in portions, $Mn_2O_3$, $Fe_2O_3$ and, finally, CuO.

In the prepared fluxes, the high-temperature phase crystallizing was ludwigite $Cu_2Mn_{1-x}Fe_xBO_5$. Single crystals of the ludwigites were synthesized by spontaneous nucleation. After homogenization of the fluxes at T = 1100°C for 3 h, the temperature was first rapidly reduced to $(T_{sat}-10)$°C and then slowly reduced with a rate of 4°C/day. In 4 days, the growth was completed, the crucible was withdrawn from the furnace, and the flux was poured out. The grown single crystals in the form of black orthogonal prisms with a length of 8 mm and a transverse size of about 1 mm were etched in a 20% water solution of nitric acid to remove the flux remainder.

Table 1. Parameters of the fluxes.

| $x$ | $p$ | $q$ | $n$, % | $T_{sat}$, °C | Designation |
|---|---|---|---|---|---|
| 0.2 | 0.6 | 0.70 | 28.0 | 905 | **S1** |
| 0.4 | 1.5 | 0.93 | 32.5 | 905 | **S2** |
| 0.5 | 1.5 | 1.12 | 36.4 | 925 | **S3** |

Three $Cu_2Mn_{1-x}Fe_xBO_5$ compounds have been synthesized. For convenience in the next parts of the present paper the samples will be indicated as S1 (sample 1), S2 (sample 2), and S3 (sample 3) – in the order of increasing of iron content – as mentioned in Table 1.

### III. Structural properties

Structural properties of the synthesized single crystals were investigated by the powder and single-crystal X-ray diffraction, and by the EXAFS/XANES techniques.

#### a. X-ray diffraction

The powder diffraction data of all of synthesized samples was collected at room temperature with a Bruker D8 ADVANCE powder diffractometer (Cu-Kα radiation) and linear VANTEC detector.

Crystal fragments of two compositions S1 and S3 were selected to the single-crystal experiment. Diffraction data were collected under room conditions using an Oxford Diffraction Xcalibur Gemini diffractometer (Mo-Kα radiation, 0.5 mm collimator, graphite monochromator) equipped with a CCD-detector. Data reduction, including a background correction and Lorentz and polarization corrections, was performed with the *CrysAlisPro* software. A semi-empirical absorption correction was applied using the multi-scan technique. The unit-cell metrics of both samples is monoclinic, space group $P2_1/c$. The structure was solved by the direct methods and refined in the anisotropic approach using SHELX-97 program package [16]. The main crystal data are shown in Table 2. The structural data are deposited as CIFs at the ICSD (CSD Nos. 433621 and 433720). The structures were refined with account taken of chemical analysis results. Atomic coordinates, occupancy (occupancy has been determined using compositions obtained by EXAFS/XANES technique (III. b)), and displacement parameters for refined structures are presented in Table 3. Selected distances and angles are listed in Table 4. It is necessary to note that the atomic X-ray scattering factors of Fe and Mn are very close to each other, so suggested distribution over atomic positions of these ions should be considered as estimated. Lattice parameters of $Cu_2Mn_{1-x}Fe_xBO_5$ ludwigites, obtained by powder and single crystal X-ray diffraction analysis in the framework of the present work in comparison with the data for pure $Cu_2MnBO_5$ (x=0) and $Cu_2FeBO_5$ (x=1) are presented in Table 5.

Table 2. The crystal structure parameters of studied compounds.

|  | S3 | S1 |
|---|---|---|
| Space group, Z | $P2_1/c$, 4 | $P2_1/c$, 4 |
| $a$ (Å) | 3.13323(5) | 3.14434(7) |
| $b$ (Å) | 12.02639(18) | 12.0255(2) |
| $c$ (Å) | 9.48739(15) | 9.46837(19) |
| $\beta$ (deg.) | 97.4539(14) | 97.530(2) |
| $V$ (Å$^3$) | 354.477(9) | 354.934(13) |
| Reflections measured/independent/with $I > 2\sigma(I)/R_{int}$ | 10719 / 1775 / 1599 / 0.0411 | 6521 / 1182 / 1118 / 0.0362 |
| $h$, $k$, $l$- limits | $-5 \leq h \leq 5$; $-20 \leq k \leq 20$; $-15 \leq l \leq 15$ | $-4 \leq h \leq 4$; $-17 \leq k \leq 17$; $-13 \leq l \leq 13$ |
| $R1 / wR2 / Goof$ for observed reflections [$I>2\sigma(I)$] | 0.0257 / 0.0595 / 1.080 | 0.0319 / 0.0855 / 1.127 |
| $R1 / wR2 / Goof$ for all data | 0.0310 / 0.0621 / 1.080 | 0.0337 / 0.0870 / 1.127 |
| $\Delta\rho_{max} / \Delta\rho_{min}$ (e/Å$^3$) | 1.588 / −1.337 | 2.182 / −1.260 |

X-ray analysis showed that ludwigites S1 and S3 are crystallizing in monoclinic-distorted structure variant of the ludwigite mineral. The studied compounds are isostructural to $Cu_2AlBO_5$ [17] and $Cu_2FeBO_5$ [18]. The cations are statistically distributed over four nonequivalent crystallographic positions. M1 and M2 are in general positions, M3 and M4 are in the special positions with symmetry -1. Analysis of the cation coordination (Table 3) shows large deviations from the octahedral symmetry that are typical for orthorhombic ludwigite structure [17]. In both studied crystals the coordination of M1 and M3 positions could be designated as [4+2]. These positions are mostly occupied by copper atoms characterized by specific distortions of the coordination polyhedra. Positions M2 and M4 are mainly occupied by iron and manganese atoms and their coordination polyhedra are less distorted. The coordination of these positions can be designated as intermediate between [4+2] and octahedral. Despite the difference of chemical compositions of studied single crystals their structure parameters are very similar. So the maximum difference between bond lengths M–O is equal 0.08 Å and observed for $M4O_6$ octahedron (see Table 4). The average difference of the bond lengths is 0.02 Å. The differences of O–M–O angles are also quite small and don't exceed one and half degrees. Obviously, it is caused by stereochemical similarity of Mn and Fe cations.

In spite of having lower symmetry of $(Cu,Fe,Mn)_3O_2BO_3$ - $P2_1/c$ - than common ludwigites have, the main motif of the structure – zig-zag walls - remains. Monoclinic distortion springs up because of copper and manganese Jahn-Teller effect [17]. Monoclinic distortion manifests itself in the orientational ordering of the long axes of the cation polyhedra (Figure 2).

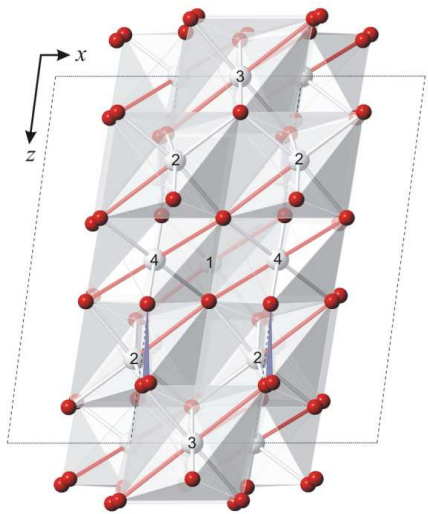

Figure 2. Projection along [010] of the $(Cu,Fe,Mn)O_2BO_3$ structure. Cation positions are indicated. Elongate interatomic bonds (M–O> 2.2 Å) are marked.

Table 3. Atomic parameters for S1 and S3 samples.

|   |   | S3 | S1 |
|---|---|---|---|
| M1 | $x$ | 0.45875(8) | 0.47026(11) |
|   | $y$ | 0.719995(19) | 0.72018(3) |
|   | $z$ | 0.00764(2) | 0.00734(3) |
|   | Occ. | $Cu_{0.8857(17)}Fe_{0.096(3)}Mn_{0.007(3)}$ | $Cu_{0.8611(17)}Fe_{0.085(3)}Mn_{0.033(3)}$ |
|   | $U_{eq}$ | 0.00809(6) | 0.00830(11) |
| M2 | $x$ | 0.93696(8) | 0.95010(12) |
|   | $y$ | 0.61804(2) | 0.61796(3) |
|   | $z$ | 0.27024(3) | 0.26936(4) |
|   | Occ. | $Cu_{0.3328(19)}Fe_{0.397(4)}Mn_{0.267(3)}$ | $Cu_{0.3704(18)}Fe_{0.161(3)}Mn_{0.460(4)}$ |
|   | $U_{eq}$ | 0.00764(7) | 0.00783(11) |
| M3 | $x$ | 0.5 | 0.5 |
|   | $y$ | 0.5 | 0.5 |
|   | $z$ | 0.5 | 0.5 |
|   | Occ. | $Cu_{0.817(3)}Fe_{0.189(4)}$ | $Cu_{0.847(3)}Fe_{0.149(4)}$ |
|   | $U_{eq}$ | 0.00720(9) | 0.00709(14) |
| M4 | $x$ | 0 | 0 |
|   | $y$ | 0.5 | 0.5 |
|   | $z$ | 0 | 0 |
|   | Occ. | $Cu_{0.401(3)}Fe_{0.362(7)}Mn_{0.248(6)}$ | $Cu_{0.450(3)}Fe_{0.119(6)}Mn_{0.444(6)}$ |
|   | $U_{eq}$ | 0.00707(9) | 0.00714(15) |
| B1 | $x$ | 0.9620(7) | 0.9691(10) |
|   | $y$ | 0.86366(16) | 0.8634(2) |
|   | $z U_{eq}$ | 0.2332(2) | 0.2332(3) |
|   |   | 0.0079(3) | 0.0087(5) |
| O1 | $x$ | 0.4589(5) | 0.4586(7) |
|   | $y$ | 0.85549(11) | 0.85593(16) |
|      | $zU_{eq}$ | –0.09796(16)   | –0.0974(2)   |
|      |           | 0.0125(3)      | 0.0122(4)    |
| O2   | $x$       | 0.9124(5)      | 0.9310(7)    |
|      | $y$       | 0.76289(11)    | 0.76315(17)  |
|      | $zU_{eq}$ | 0.16447(15)    | 0.1638(2)    |
|      |           | 0.0112(2)      | 0.0136(4)    |
| O3   | $x$       | 0.0351(4)      | 0.0392(7)    |
|      | $y$       | 0.46041(11)    | 0.46066(17)  |
|      | $zU_{eq}$ | 0.34334(15)    | 0.3427(2)    |
|      |           | 0.0096(2)      | 0.0109(4)    |
| O4   | $x$       | 0.4857(7)      | 0.5161(10)   |
|      | $y$       | 0.57854(15)    | 0.5790(2)    |
|      | $zU_{eq}$ | 0.1137(2)      | 0.1117(3)    |
|      |           | 0.0342(5)      | 0.0328(7)    |
| O5   | $x$       | 0.0103(5)      | 0.0163(7)    |
|      | $y$       | 0.63564(11)    | 0.63636(15)  |
|      | $zU_{eq}$ | –0.12062(15)   | –0.1204(2)   |
|      |           | 0.0106(2)      | 0.0105(4)    |

Table 4. Selected distances (Å) and angles (°) for S1 and S3 ludwigites.

|              | S3          | S1          |
|--------------|-------------|-------------|
| M1–O1        | 1.9130(14)  | 1.908(2)    |
| M1–O2        | 1.9879(14)  | 2.000(2)    |
| M1–O2        | 2.464(2)    | 2.449(2)    |
| M1–O4        | 1.9725(17)  | 1.960(2)    |
| M1–O5        | 2.0102(14)  | 2.016(2)    |
| M1–O5        | 2.458(2)    | 2.444(2)    |
| Mean         | 2.134       | 2.130       |
|              |             |             |
| M2–O1        | 1.9516(15)  | 1.928(2)    |
| M2–O1        | 2.0963(17)  | 2.143(2)    |
| M2–O2        | 2.0069(14)  | 2.009(2)    |
| M2–O3        | 2.0283(14)  | 2.022(2)    |
| M2–O4        | 1.9714(17)  | 1.944(3)    |
| M2–O4        | 2.460(3)    | 2.512(4)    |
| Mean         | 2.086       | 2.093       |
|              |             |             |
| M3–O1(2×)    | 1.9671(14)  | 1.959(2)    |
| M3–O3(2×)    | 1.9989(13)  | 1.994(2)    |
| M3–O3(2×)    | 2.4277(14)  | 2.445(3)    |
| Mean         | 2.131       | 2.133       |
|              |             |             |
| M4–O4(2×)    | 1.9863(17)  | 2.050(3)    |
| M4–O4(2×)    | 2.259(3)    | 2.181(3)    |
| M4–O5(2×)    | 1.9953(14)  | 2.0016(19)  |
| Mean         | 2.080       | 2.078       |
|              |             |             |
| B1–O2        | 1.375(2)    | 1.370(4)    |
| B1–O3        | 1.372(2)    | 1.372(3)    |
| B1–O5        | 1.375(2)    | 1.375(4)    |
|              |             |             |
| Mean         | 1.374       | 1.372       |

Table 5. Lattice parameters of $Cu_2Mn_{1-x}Fe_xBO_5$ ludwigites, obtained by powder and single crystal X-ray diffraction analysis in the framework of the present work in comparison with the data for pure $Cu_2MnBO_5$ (x=0) and $Cu_2FeBO_5$ (x=1).

|         | a, Å       | b, Å       | c, Å      | β        | Ref.     |
|---------|------------|------------|-----------|----------|----------|
| x=1     | 3.108      | 12.003     | 9.459     | 96.66°   | [15]     |
| x=0.5 (S3) | 3.1339(3) | 12.0204(1) | 9.4855(5) | 97.477°  | pr. work |
| x=0.4 (S2) | 31360(3)  | 12.0178(2) | 9.4865(6) | 97.549°  | pr. work |
| x=0.2 (S1) | 3.1443(1) | 12.0255(2) | 9.4684(2) | 97.53°   | pr. work |
| x=0     | 3.14003    | 12.0242    | 9.3973    | 92.261°  | [12]     |

### b. XANES/EXAFS

XANES and EXAFS spectra at the Fe, Mn, and Cu *K*-edges were recorded at room temperature in the transmission mode at the Structural Materials Science beamline of the Kurchatov Synchrotron Radiation Source (National Research Center "Kurchatov Institute", Moscow) [19]. For the selection of the beam photon energy, a Si (111) channel-cut monochromator was employed, that provided an energy resolution was $\Delta E/E \sim 2 \cdot 10^{-4}$. Incident and transmitted intensities were recorded using two ionization chambers filled with appropriate $N_2$/Ar mixtures to provide 20% and 80% absorption.

The energies were calibrated against a sharp pre-edge feature of $KMnO_4$ (Mn *K*-edge) as well as using Fe and Cu metal foils (Fe and Cu *K*-edges, respectively). The EXAFS spectra were collected using optimized scan parameters of the beamline software. The *ΔE* scanning step in the XANES region was about 0.45 eV, and scanning in the EXAFS region was carried out at a constant step on the photoelectron wave number scale with $\Delta k = 0.05$ Å$^{-1}$ that corresponds to the energy step of the order of 1.5 eV. The signal integration time was 4 s/point. Single-crystalline $Cu_2Mn_{1-x}Fe_xBO_5$ samples were ground to fine powders and then spread uniformly onto a thin adhesive Kapton film which was folded several times to provide an absorption edge jump around unity.

The EXAFS spectra *μ*(*E*) were normalized to a unit edge jump and the isolated atom absorption coefficient $\mu_0(E)$ was extracted by fitting a cubic-spline-function versus the experimental data. After subtraction of the smooth atomic background, the conversion from photon energy *E* to photoelectron wave number *k* scale was performed. Crystallographic parameters were used as a starting structural model. The $k^3$-weighted EXAFS function $\chi_{exp}(k)$ was calculated in the intervals $k = 2 - 13$ Å$^{-1}$ using a Kaiser-Bessel window. The EXAFS structural analysis was performed using theoretical phases and amplitudes as calculated by the FEFF package [20], and fits to the experimental data were carried out in the *R*-space with the IFFEFIT package [21].

Table 6. Jumps of the absorption coefficients and determined chemical compositions of $Cu_2Mn_{1-x}Fe_xBO_5$ ludwigites.

|                  | S1                              | S2                              | S3                              |
|------------------|---------------------------------|---------------------------------|---------------------------------|
| x (in flux)      | 0.2                             | 0.4                             | 0.5                             |
| Fe-K             | 0.5989                          | 0.4721                          | 0.6641                          |
| Cu-K             | 2.2830                          | 0.9787                          | 1.2207                          |
| Mn-K             | 1.2720                          | 0.3186                          | 0.3753                          |
| Real composition | $Cu_{1.88}Mn_{0.74}Fe_{0.38}BO_5$ | $Cu_{1.87}Mn_{0.43}Fe_{0.7}BO_5$ | $Cu_{1.83}Mn_{0.40}Fe_{0.77}BO_5$ |

Normalized XANES spectra on the *K*-edges of 3*d* metals at T=300 K of $Cu_2Mn_{1-x}Fe_xBO_5$ ludwigites are shown in Figure 3. Due to the high sensitivity of the spectra to atoms of the studied compounds, it is possible to determine the weight content of 3*d* ions by the rations of the main peaks intensities on *K*-edges of Fe, Mn and Cu. The results of the composition are in Table 6. The results are shown in Table 6.

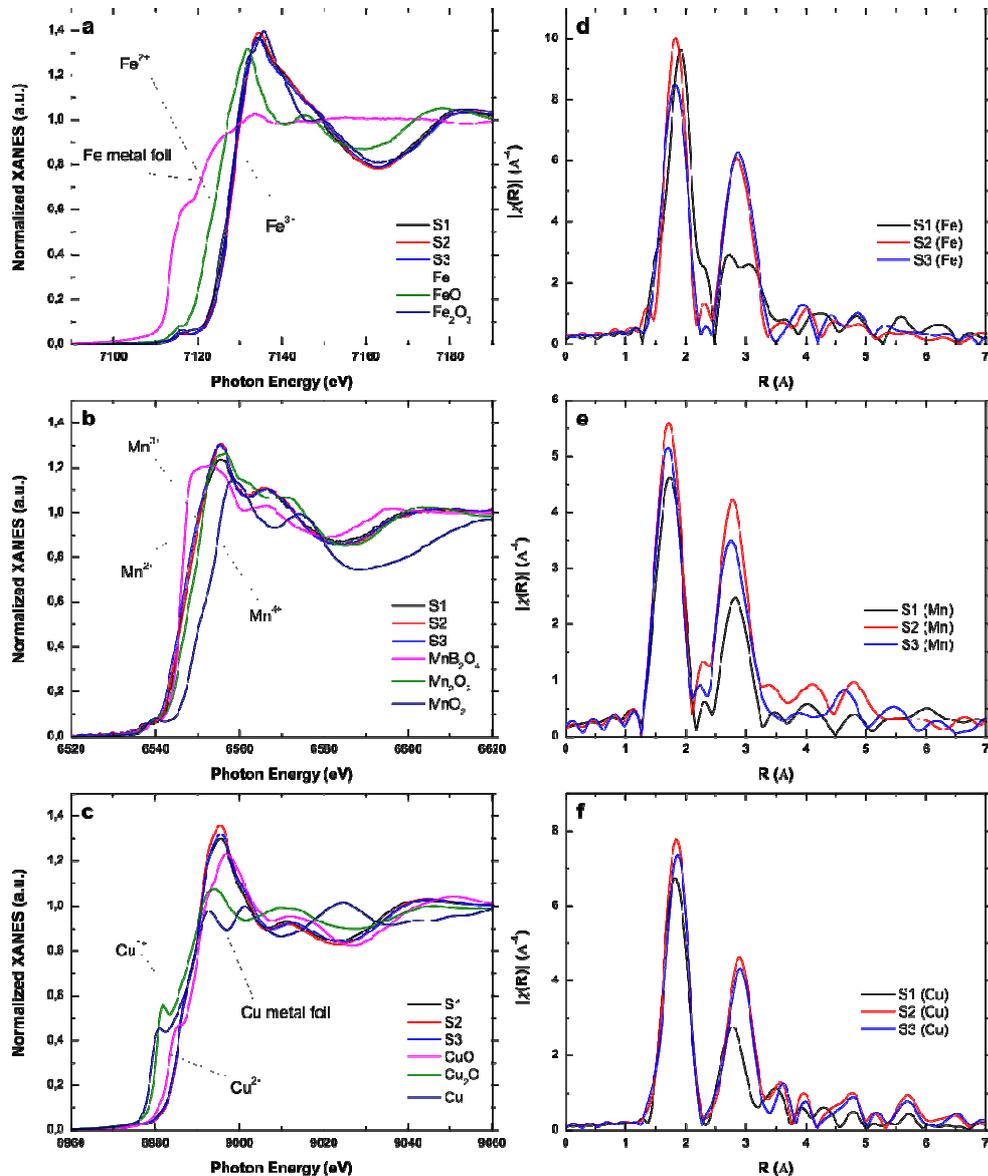

Figure 3. Normalized XANES spectra (a, b, c) and Fourier-transforms (d, e, f) of $k^3$-weighted EXAFS spectra on *K*-edges of Fe (a, d), Mn (b, e) and Cu (c, f) at T=300 K of S1, S2 and S3. The XANES spectra of standards ($Fe^{2+,3+}$, $Mn^{2+,3+,4+}$, and $Cu^{1+,2+}$) are shown for a comparison of charge states of 3*d* atoms.

In the range of *K*-edge absorption it is possible to highlight three features which could be interpreted as transitions of photoelectron excited from the 1*s*-level of Fe, Mn and Cu to bound states and the processes of its scattering on the local environment. The features located previous to the main absorption edge corresponds to 1*s* - 3*d* quadrupole transition for octahedra coordination (~7118 eV, ~6540 eV for Fe and Mn, respectively). The weak intensity of the pre-edge peak indicates a slightly distorted six-coordinated octahedral environment around the 3*d*-atoms as it is expected for the ludwigite structure. The main edge positions observed for the $Cu_2Mn_{1-x}Fe_xBO_5$ ludwigites are the same as that for $Fe_2O_3$, $Mn_2O_3$ and CuO standards, clearly indicating the $Fe^{3+}$, $Mn^{3+}$ and $Cu^{2+}$ states for the S1, S2 and S3 samples, respectively. The main absorption maximum corresponds to dipole-resolved $1s$ - $4p$ transition. The peculiarities of the fine structure of the upper energy range have EXAFS origin.

Forms of Fourier-transforms of $Cu_2Mn_{1-x}Fe_xBO_5$ ludwigites (Figure 3) at the $K$-edges of Cu, Fe and Mn are quite complex. It related to the distortions of the oxygen octahedra $CuO_6$, $MnO_6$ and $FeO_6$. As it shown in Figure 3 the modules of Fourier-transforms of EXAFS spectra consist of the first peaks ($r \approx 1.0$ - $2.1$ Å) corresponding to Me – O coordination spheres, the second peaks ($r \approx 2.20$ - $3.10$ Å) corresponding to Me – Me coordination spheres and peaks of less intensity, which corresponds to effects of multiple scattering, far coordination spheres Me – Me and Me – B distances. In order to receive local crystal parameters the XRD structural data of $Cu_2Mn_{1-x}Fe_xBO_5$ ludwigites were adopted to calculate theoretical amplitude and phases for each scattering path up to 6 A. Six interatomic distances $R_{Me-O}$ with common Debay-Waller factor $\sigma^2$ were varied to obtain best fits. The main results of the structural analysis at Fe, Mn and Cu $K$-edges are summarized in Table 7. The average interatomic distances <Fe-O> slightly decrease with increasing Fe content. In turn, the average interatomic distances <Mn-O> slightly inrease with decreasing Mn content. This can result from the mutual influences of local strains in $FeO_6$ and $MnO_6$ octahedra. Comparing the Me-O average distances obtained from the EXAFS and XRD data, we can conclude that Cu ions are in the crystallographic positions of M1 and M3, preferentially, while Fe and Mn are mainly in positions M2 and M4. This fact allows us to broaden our understanding of the population of crystallographic positions by metal ions.

Table 7. Best fit structural parameters of the first oxygen coordination shell for $Cu_2Mn_{1-x}Fe_xBO_5$ ludwigites at the Fe, Cu, and Mn $K$-edges, where $N$ is the coordination number, $R$ is the inreatomic distances for the octahedral site, $\sigma^2$ are Debay-Waller factors and $R_f$ is the fitting discrepancy factor. The average interatomic distance $<R_{Me-O}>$ is highlighted bold.

| | $N$ | $R_{Fe-O}$ (Å) | $\sigma^2_{Fe-O}$ (Å$^2$) | $R_{Fe-O}$ (%) | $N$ | $R_{Mn-O}$ (Å) | $\sigma^2_{Mn-O}$ (Å$^2$) | $R_{Mn-O}$ (%) | $N$ | $R_{Cu-O}$ (Å) | $\sigma^2_{Cu-O}$ (Å$^2$) | $R_{Cu-O}$ (%) |
|---|---|---|---|---|---|---|---|---|---|---|---|---|
| | 1.20 | 1.91(2) | | | 0.83 | 2.08(2) | | 5 | 0.85 | 2.11(2) | | 3.6 |
| | 1.20 | 2.04(2) | | | 0.83 | 1.94(2) | | | 0.85 | 1.94(2) | | |
| | 1.20 | 2.06(2) | | | 0.83 | 1.94(2) | 3.97 ·10$^{-3}$ | | 0.85 | 2.01(2) | 1.77 ·10$^{-3}$ | |
| **S1** | 1.20 | 2.06(2) | 1·10$^{-3}$ | 1 | 0.83 | 1.87(2) | | | 0.85 | 1.93(2) | | |
| | 1.20 | 2.27(2) | | | 0.83 | 2.20(2) | | | 0.85 | 2.36(2) | | |
| | 1.20 | 2.42(2) | | | 0.83 | 2.37(2) | | | 0.85 | 2.55(2) | | |
| | | **2.13(2)** | | | | **2.07(2)** | | | | **2.15(2)** | | |
| | 1 | 2.07(2) | | | 1 | 1.93(2) | | 2.4 | 0.85 | 1.93(2) | | 1.5 |
| | 1 | 2.02(2) | | | 1 | 1.92(2) | | | 0.85 | 1.93(2) | | |
| | 1 | 1.95(2) | 0.65 ·10$^{-3}$ | | 1 | 1.83(2) | 3.02 ·10$^{-3}$ | | 0.85 | 2.07(2) | 0.4 ·10$^{-3}$ | |
| **S2** | 1 | 1.95(2) | | 1.7 | 1 | 2.08(2) | | | 0.85 | 2.04(2) | | |
| | 1 | 2.17(2) | | | 1 | 2.27(2) | | | 0.85 | 2.30(2) | | |
| | 1 | 2.35(2) | | | 1 | 2.54(2) | | | 0.85 | 2.47(2) | | |
| | | **2.08(2)** | | | | **2.09(2)** | | | | **2.12(2)** | | |
| | 1 | 2.10(2) | | | 1 | 1.92(2) | | 1.3 | 0.765 | 1.94(2) | | 1 |
| | 1 | 1.96(2) | | | 1 | 2.08(2) | | | 0.765 | 2.02(2) | | |
| | 1 | 1.96(2) | 3.63 ·10$^{-3}$ | | 1 | 1.91(2) | 5.56 ·10$^{-3}$ | | 0.765 | 2.02(2) | 3.19 ·10$^{-3}$ | |
| **S3** | 1 | 1.96(2) | | 1 | 1 | 2.22(2) | | | 0.765 | 1.94(2) | | |
| | 1 | 2.08(2) | | | 1 | 1.87(2) | | | 0.765 | 2.34(2) | | |
| | 1 | 2.35(2) | | | 1 | 2.51(2) | | | 0.765 | 2.52(2) | | |
| | | **2.07(2)** | | | | **2.09(2)** | | | | **2.13(2)** | | |

## IV. Magnetic properties

Magnetic measurements of single crystals $Cu_2Mn_{1-x}Fe_xBO_5$ were performed using the physical properties measurements system PPMS-9 (Quantum Design) at temperature range $T=3 \div 300$ K and magnetic fields up to 90 kOe.

Experimental study of the magnetic structure of $Cu_2MnBO_5$ ludwigite by NPD revealed the absence of the ordering of the magnetic moments of copper on 2a crystallographic position. That could be caused by weak exchange coupling between copper and neighbor ions [10]. However, unlike $Cu_2FeBO_5$ ludwigite, in this compound the manganese cations occupy only one crystallographic position 4e. The iron cations in $Cu_2FeBO_5$ ludwigite could be located at different crystallographic positions as it was shown by Mössbauer effect [14, 15]. It means that in substituted $Cu_2Mn_{1-x}Fe_xBO_5$ ludwigites, part of iron cations could occupy the same positions as copper and known magnetic structure of parent compound ($Cu_2MnBO_5$ ludwigite) will change to a large extent.

To estimate of the influence of $Mn^{3+} \rightarrow Fe^{3+}$ substitutions to magnetic structure the magnetic properties of three compounds $Cu_2Mn_{1-x}Fe_xBO_5$ (S1, S2, S3, as it was indicated in the Chapter II) synthesized in frameworks of the present work were studied. The results have been divided by two parts: magnetic properties of S1 and magnetic properties of samples S2 and S3.

### a. Magnetic properties of S1 sample

Thermal-field dependencies of magnetization of $Cu_{1.88}Mn_{0.74}Fe_{0.38}BO_5$ (S1) ludwigite obtained at the orientations of external magnetic field H∥a, H⊥a with value H=1 kOe are shown in Figure 4. In agreement with the presented dependencies of the magnetization it is clearly seen that S1 undergo magnetic phase transition at the temperature range T≈40÷50 K marked by the magnetization increasing. Below the phase transition temperature there are broad peak of the magnetization and further weak decreasing of the magnetization at low temperatures in FC regime (cooling of the sample in nonzero magnetic field). The measurements of the magnetization at different magnetic fields shown that this broad peak is observed up to the magnetic field H=5 kOe (Figure S1). At the magnetic field H=5 kOe and higher the low-temperature magnetization demonstrates a weak dependence on the temperature in FC regime. Difference of FC and ZFC curves at low temperatures depends on the value of applied magnetic field and could be associated with both domain walls movement and presence of spin-glass state.

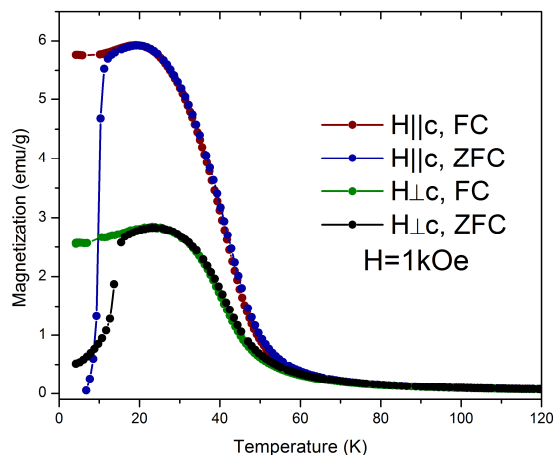

Figure 4. Temperature dependence of the magnetization of S1 (H∥a, H⊥a, H=1 kOe).

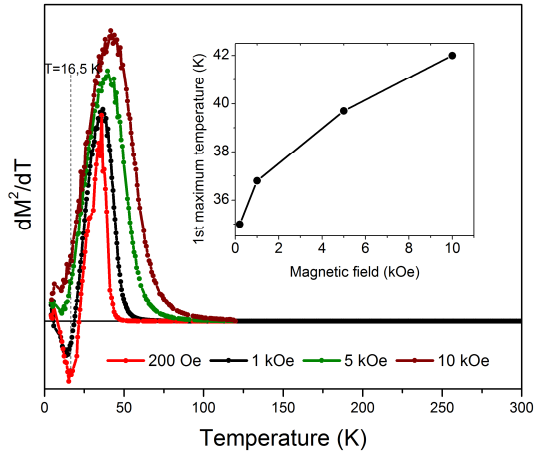

Figure 5. Temperature dependences of the temperature derivative of the squared magnetization $dM^2/dT(T)$ of S1 obtained at H||a (FC regime). Inset: dependence of the temperature of the high-temperature maximum of the derivative on the applied magnetic field.

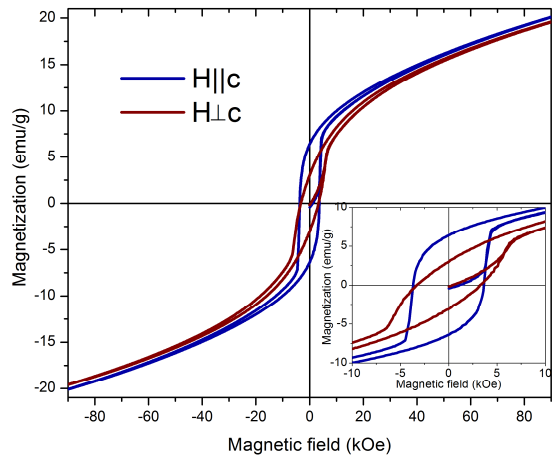

Figure 6. Magnetic field dependencies of the magnetization of S1 obtained at T=4.2 K, H||a, H⊥a.

The value of the magnetic moment of the sample S1 is different for different directions of the applied magnetic field and that designates the anisotropy of the magnetization in this sample (Figure 4). The magnetic moment along the *a* axis is almost twice larger than the magnetic moment measured along the *bc* plane. This observation qualitatively agrees with the behavior of unsubstituted $Cu_2MnBO_5$ ludwigite. However, the value of the magnetic moment along *a* axis for S1 is four times less than for parent compound. This could indirectly indicate the reducing of the degree of the magnetic ordering.

For the exact determination of the temperature of the magnetic phase transition the temperature dependences of the temperature derivative of the squared magnetization $dM^2/dT(T)$ have been built (Figure 5). Since, according to the molecular field theory, the magnetic contribution to the specific heat is proportional to the squared spontaneous magnetization [22]. It is necessary to point out that there are two extremums on the $dM^2/dT(T)$ curves: the first one corresponds to early mentioned magnetic phase transition, and the other one, of small amplitude, takes place at low temperatures. The analysis of the curves showed the dependence of the temperatures of the centers of these anomalies on the applied magnetic field value (inset in Figure 5 for the first phase transition). In magnetic fields of 0.2÷10 kOe the temperature of the first magnetic phase transition change in the range $T_1$=36÷42 K; the temperature of the second, low temperature extremum, change in the range $T_2$=11÷16.5 K. These temperatures are significantly lower than the ordering temperature of parent $Cu_2MnBO_5$ ludwigite compound ($T_c$=92 K [10]). The lowering of the temperature of the magnetic phase transition could indicate the increasing the exchange interaction competing and this is typical for Mn-Fe compounds.

Field dependencies of the magnetization of single crystal S1 are presented in Figure 6. These dependencies have been obtained at temperature T=4.2 K and the orientation of the applied magnetic field was H||a, H⊥a. Magnetic hysteresis is observed for both orientations of the sample, but the value of the magnetic moment is significantly different and it is in agreement with thermal dependencies of the magnetization. In an available range of the magnetic fields (up to 9 T) the hysteresis loops are minor (unsaturated hysteresis loops) for both directions of the applied field.

### b. Magnetic properties of S2 and S3 samples

Temperature dependencies of magnetization of $Cu_{1.87}Mn_{0.43}Fe_{0.7}BO_5$ (S2) and $Cu_{1.83}Mn_{0.40}Fe_{0.77}BO_5$ (S3) ludwigites are presented in Figure 7. These dependencies have been obtained at the orientations of applied magnetic field H||a, H⊥a of value H=1 kOe. Magnetic behavior of S2 and S3 is significantly different from the sample S1: there is a smooth increasing of the magnetization in paramagnetic phase and at the temperature T≈27 K (for both S2 and S3) the FC dependencies demonstrate "a bend" associated with the magnetic phase transition. At low temperature a slight increasing of magnetization is observed in FC regime. ZFC curves demonstrate a maximum at the temperature of the phase transition. The width of this maximum increases as the magnetic field value increases. In comparison with the sample S1 containing small iron substitution the value of magnetic moments of S2 and S3 has decreased by two orders of magnitude.

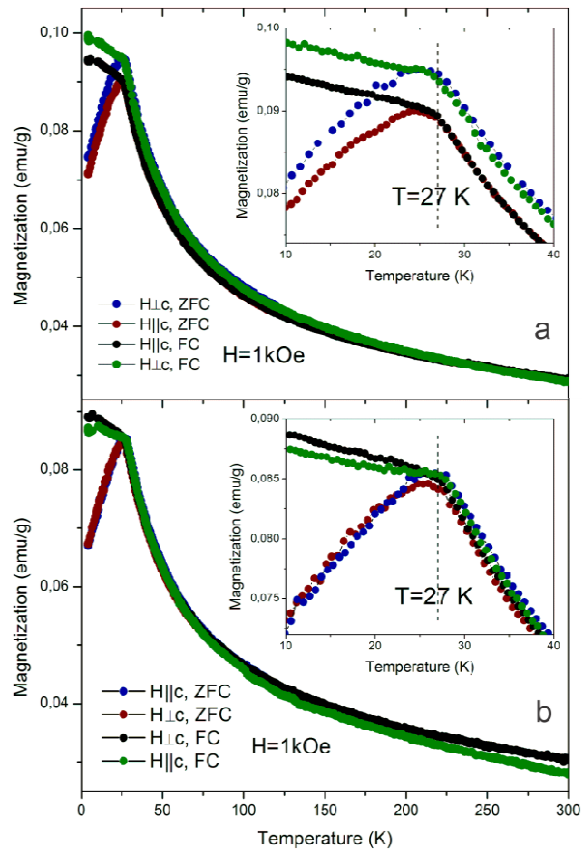

Figure 7. Temperature dependence of magnetization of S2 (a) and S3 (b) (H||a, H⊥a, H=1 kOe).

Field dependencies of the magnetization of single crystals S2 and S3 are presented in Figure 8. These dependencies have been obtained at temperature T=4.2 K and the orientation of the applied magnetic field was H||a, H⊥a. Both samples demonstrate minor (unsaturated) hysteresis loops in available range of the magnetic fields (up to 9 T) for both directions of applied magnetic field. Despite the high similarity of the magnetic behavior the sample S3 (with larger iron content) has significant features: field dependencies of the magnetization are anisotropic unlike the sample S2. This could indicate the increasing of the magnetic ordering degree and restoration of magnetic order. However the value of magnetization at H=9 T of S3 single crystal is less than the value of magnetization of S2 sample for both directions despite the increasing of iron content.

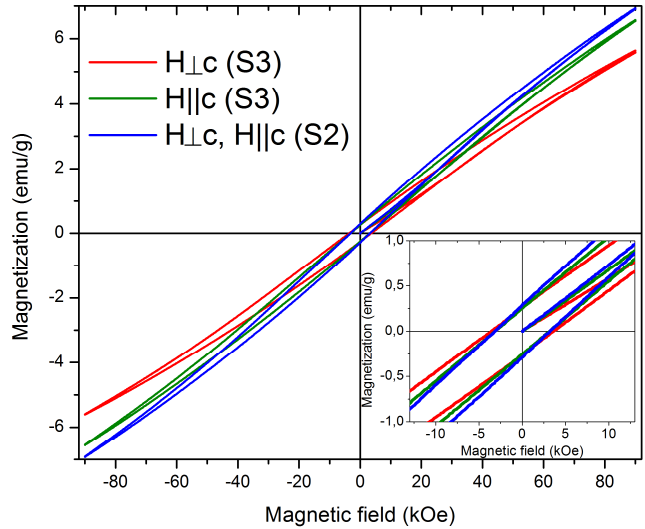

Figure 8. Magnetic field dependencies of the magnetization of S2 and S3 single crystals obtained at T=4.2 K, H||a, H⊥a.

## V. Discussion

In this work we report the synthesis and characterization of the system $Cu_2Mn_{1-x}Fe_xBO_5$ using X-ray diffraction, XANES/EXAFS and magnetization measurements. In the process of flux growth of single crystals it was supposed to substitute the cations $Mn^{3+}$ by the $Fe^{3+}$ cation in the parent compound $Cu_2MnBO_5$. But the structure characterization of synthesized samples showed the presence of manganese cations in a divalent subsystem of copper. The magnitude of such substitution is not high about 6-8% depending on the sample. Along with the presence of the manganese cations in the divalent subsystem the total amount of iron content is exceed from the initial data in fluxes. The question on the origin of such concentration discrepancy is still opened and requires additional synthesize experiments. But the most probable explanation of these results is distinction of the distribution coefficients of $Cu^{2+}$, $Mn^{3+}$ and $Fe^{3+}$ at the crystal formation in used fluxes. High solubility of CuO in these fluxes leads to small distribution coefficient of $Cu^{2+}$ in the crystal and, as a result of lack of copper, the crystal contains cations of divalent manganese. The solubility of $Mn_2O_3$ and $Fe_2O_3$ in used fluxes is similar, but there is small predominance of manganese coefficient, so the total amount of iron is slightly larger. Similarity of the solubility of $Mn_2O_3$ and $Fe_2O_3$ in the used fluxes is confirmed by small concentration deviations from the initial flux content for the system $Mn_{2-x}Fe_xBO_4$ with warwickite structure (it was used the same solvent) [23]. The suggestion based on the solubility of the oxides is proved by the real cation concentration in the synthesized single crystals.

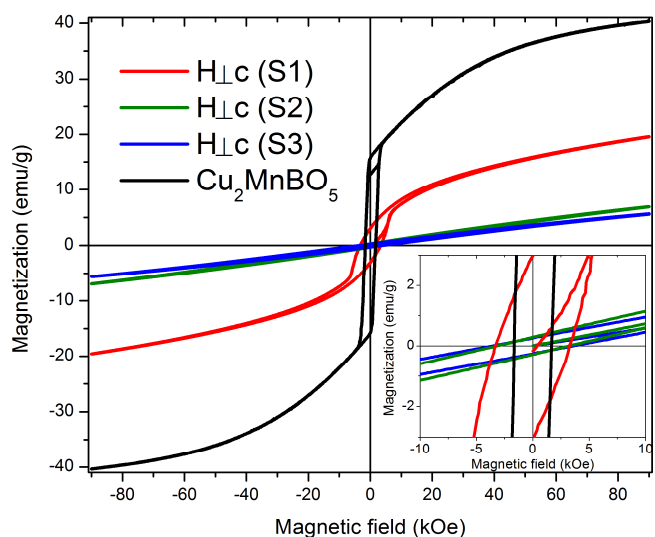

Figure 9. Magnetic field dependencies of the magnetization of S1, S2, S3 and parent $Cu_2MnBO_5$ single crystals obtained at T=4.2 K, H⊥a.

For better understanding of the evolution of the magnetic structure of Cu-Mn ludwigite under iron addition, the field dependencies of magnetization for three synthesized samples and parent compound $Cu_2MnBO_5$ are shown in Figure 9. These dependencies have been obtained at the temperature T=4.2 K, the magnetic field was applied along *a* axis. It is clear that unsubstituted ludwigite $Cu_2MnBO_5$ has a maximum magnetic moment in comparison with others for all range magnetic fields. This sample has a closed hysteresis loop close to a square shape (with vertical walls). The samples S2 and S3 (with comparable iron and manganese content) are characterized by the lowering of magnetic moment by the order of magnitude, and the shape of the hysteresis loop is fully changed: the loops became opened in magnetic fields up to 90 kOe. Such behavior of the field dependency of magnetization is caused by nonequilibrium ($M_{FC}(H)$) state of the system [24]. The field dependence of magnetization of the sample S1 (with less addition of the iron) is intermediate between the analogy dependence of the parent Cu-Mn ludwigite and the ludwigites with the comparable Mn/Fe ratio: the value of the magnetic moment is only twice less than in parent compound, the shape of the field dependency can be decomposed into two hysteresis loops. The first one is opened hysteresis loop typical for the samples S2 and S3, the second one is closed hysteresis loop typical for unsubstituted ludwigite. And the value of $H_c$ of the sample S1 has increased in comparison with parent compound ($H_c$=1.7 kOe for $Cu_2MnBO_5$) and it equals to $H_c$=3.3 kOe. So it could be concluded that in ludwigite S1 there are a coexistence of two magnetic phases. The value of the coercive field $H_c$ in S1 corresponds to the value of this parameter for S2 and S3, that also confirms the hypothesis on the relationship of the magnetic phases of S1 and S2 and S3 (Figure 9).

As it was mentioned in the previous work [10], the experimental studying of the magnetic structure of unsubstituted $Cu_2MnBO_5$ ludwigite by NPD has shown the low magnetic moment of copper on the *2a* site, that could indicate the partial magnetic disordering of this site at low temperature phase. In the case of Fe-substituted ludwigites, from the structural point of view the part of the disordered positions increases. And as a consequence of structural disorder and strong magnetic frustrations caused by the presence of the Fe and Mn cations it should lead to increasing of the magnetic disorder and lowering of the total magnetic moment. From the experiment one can observe the dramatic lowering of the moment from parent compound to the samples S2 and S3 which coincides to the suggestion of the magnetic disordering. Taking into account the strong exchange interactions competition it was suggested a hypothesis about the spin-glass state origin of the magnetic phase realized in S2, S3 and partially in S1 samples. To verify this hypothesis the measurements of the *ac*-susceptibility of S1, S2 and S3 samples have been carried out. The results of these measurements are presented in Figure 10.

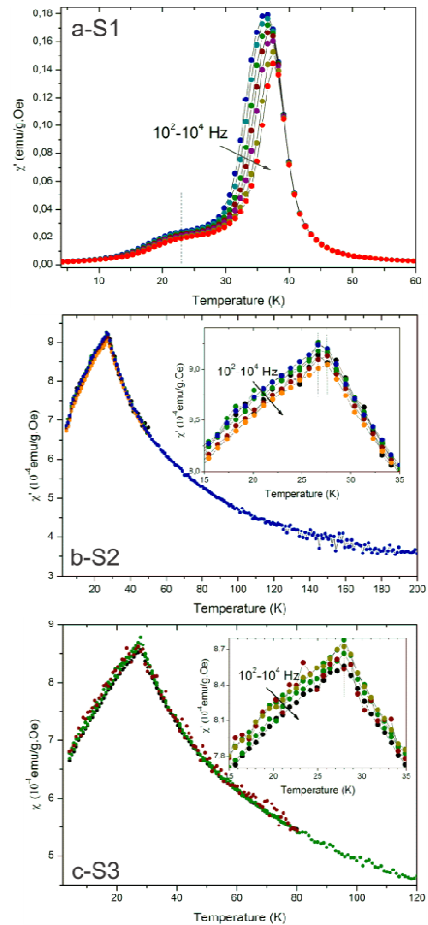

Figure 10. Real part of S1 (a), S2 (b) and S3 (c) *ac*-magnetic susceptibility as functions of temperature. The amplitude of the oscillating magnetic field is 10 Oe.

The *ac*-susceptibility measurements (Figure 10) of the synthesized samples S1, S2 and S3 showed distinctly different behavior: the frequency dependence of the temperature of the magnetic phase transition is present only for S1 and S2 that could indicate spin-glass state in the low temperature phase. The sample S3 does not show any frequency dependence, which implies that the magnetic phase transition in S3 is not due to freezing effects as in glassy systems. Besides the frequency dependence it was obtained that in the sample S1 there are two magnetic phase transitions: the first one – early identified – at the temperature range $T_1=33.5 \div 36$ K ($\Delta T=2.5$ K at frequency of the external magnetic field $10^2 \div 10^4$ Hz), the second one – at the temperature range $T_2=20.4 \div 21.5$ K ($\Delta T=1.1$ K at frequency of the external magnetic field $10^2 \div 10^4$ Hz). So, both magnetic phase transitions in S1 show frequency dependence denoted the spin-glass state presence.

Probably the shape of hysteresis loop of the sample S1 at low temperatures – two nested loops – is the result of the consequent phase transitions. Due to the complexity of the crystallographic and magnetic structure and the presence of four nonequivalent magnetic cation positions in the unit cell each of these two magnetic phase transitions can be related to distinct magnetic subsystem. So the magnetic behavior of the sample S1 can be compared with

Cu$_2$FeBO$_5$ which demonstrates three sequential magnetic phase transitions including the first one to spin-glass state of only iron subsystem [14].

The sample S2 doesn't have such strong frequency dependence as S1: the temperature difference is ΔT=1 K at the frequency of the external magnetic field $10^2 \div 10^4$ Hz. However the field dependencies of the magnetization of this sample is totally different from the parent sample Cu$_2$MnBO$_5$, and there are no anisotropy. Of course, along with the disordering and possible freezing of the magnetic moments (spin glass state) the shape of M(H) dependency of S2 and S3 can also indicate the increasing of antiferromagnetic interactions in the crystal. But the absence of the anisotropy in S2 for ludwigite structure can't accord with the antiferromagnetic state. Due to this the authors are inclined to believe that the sample S2 at low temperatures has either the static disordered state (freezing of the magnetic moments at T$_c$) or partially static disordered state together with partially dynamic disordered state (paramagnetic) of some part of magnetic sites.

The analysis of the temperature dependencies of *ac*-susceptibility of S3 showed that this sample doesn't reveal the frequency dependence of the transition temperature. So, the time dependent effects typical for spin-glass state is absent in this sample. The appearance of the anisotropy of the field dependencies of the magnetization also indicates the recovering of the magnetic order in the crystal. However, the hysteresis loops of S3 are opened as for others samples that is not typical for antiferromagnets. So it is suggested that despite the absence of the frequency dependency of the transition temperature and the presence of the anisotropy of M(H) dependencies, the fully ordered magnetic state isn't realized in the sample S3. And at the magnetic phase transition there is ordering of only part of the magnetic cation sites, the others remain as paramagnetic disordered.

Table 8. The Curie-Weiss temperatures and the temperatures T$_c$ of the magnetic phase transitions of the samples S1, S2, S3 and parent compound Cu$_2$MnBO$_5$ obtained for different orientations of the magnetic field (H∥a and H⊥a) via fitting of the temperature dependencies of the reversal molar magnetic susceptibility by the Curie-Weiss law.

|  | Cu$_2$MnBO$_5$ | S1 | S2 | S3 |
|---|---|---|---|---|
| θ$_{CW}$ (H∥a), K | 50.9 | -100.2 | -338.1 | -363.0 |
| θ$_{CW}$ (H⊥a), K | 73.6 | -61.5 | -257.2 | -241.7 |
| T$_c$, K | 92 | 36 | 27 | 27 |

To analyze the temperature dependencies of the reversal molar magnetic susceptibility the Curie-Weiss law has been used [25]. The experimental curves of the reversal magnetic susceptibility have been fitted in the temperature range far from the phase transition temperatures. The Curie-Weiss temperatures of the samples S1, S2, S3 and parent compound Cu$_2$MnBO$_5$ obtained for different orientations of the magnetic field are presented in Table 8.

As a result of the fitting, it is obvious that the addition of the iron increases the antiferromagnetic interactions. The large negative paramagnetic temperatures of the samples S2 and S3 (Table 8) indicate the predominance of antiferromagnetic couplings. The Fe-induced cation disorder superimposed on the triangle network of the magnetic moments gives rise to enhance of the magnetic frustration role that is reflected in the large ratio |θ$_{CW}$|/T$_c$>10 [23] for the samples S2 and S3. The Curie-Weiss temperatures for different directions of the magnetic field are different, and it is consequence of the anisotropy of the magnetic susceptibility in paramagnetic phase. The same effect has been observed for parent compound Cu$_2$MnBO$_5$ and related to anisotropy of the g-factor due to the Jahn-Teller effect of Cu$^{2+}$ and Mn$^{3+}$ [10].

## VI. Conclusions

Oxyborates of the ludwigite type are wide class of compounds with rich magnetic behavior. The main features of these compounds are caused by the structure: quasi-low-dimensionality, mixed valence of the magnetic cations, triangle network, the presence of four nonequivalent positions of magnetic cations. Ludwigites $Cu_2Mn_{1-x}Fe_xBO_5$ studied in this work are bright representatives of this family. The magnetic behavior study of these ludwigites has shown the high sensitivity of the magnetic properties even to small composition variation.

Three samples of $Cu_2Mn_{1-x}Fe_xBO_5$ ludwigites with x=0.2, 0.4, 0.5 have been synthesized by flux method. The obtained single crystals were characterized from structural and magnetic point of view. It was confirmed the monoclinic distorted ludwigite structure with $P2_1/c$ space group and the phase homogeneity for all the samples. Lattice parameters, bond lengths, site occupancies have been investigated using X-ray diffraction techniques. Using XANES/EXAFS the composition $x$ of the synthesized single crystals as well as the local structure of the transitional metal cations have been studied. The exact chemical formulas have been obtained: $Cu_{1.88}Mn_{0.74}Fe_{0.38}BO_5$ (S1), $Cu_{1.87}Mn_{0.43}Fe_{0.7}BO_5$ (S2), $Cu_{1.83}Mn_{0.40}Fe_{0.77}BO_5$ (S3). The charge state of the Cu, Fe, Mn ions has been determined. Magnetic properties studying of the synthesized samples have helped to estimate the way of the Fe-addition influence to the parent compound $Cu_2MnBO_5$. It was found that the different magnetization behavior for each studied samples that emphasize the high sensitivity of the magnetic properties of ludwigites to small composition variation. The main point of the investigation of the magnetic properties was the magnetic ordering degree. It was established that as $x$ increasing up to $x$=0.4 the magnetic order of $Cu_2Mn_{1-x}Fe_xBO_5$ ludwigites was destroyed. But as the continuing of the $x$ increasing there is recovering of the magnetic order. We hope that this study will help in the understanding of the processes in ludwigites and Mn-Fe compounds with the other structure.

**Acknowledgements**
This study was supported by Russian Foundation for Basic Research (RFBR) and Government of Krasnoyarsk Territory according to the research project No. 16-42-243028. X-ray diffraction experiment was carried out at the Centre on Molecular Design and Ecologically Safe Technologies at the Novosibirsk State University. The work of one of the coauthors (M.S.P.) was supported by the Russian Foundation for Basic Research (project no. 16-32-60049) and by the program of Foundation for promoting the development of small enterprises in scientificand technical sphere ("UMNIK" program).